\documentclass[runningheads]{llncs}
\pdfoutput=1
\usepackage{amsmath}
\usepackage{array}
\usepackage[dvipsnames]{xcolor}
\usepackage{footmisc}
\usepackage{graphicx}
\usepackage{multirow}
\usepackage{url}
\usepackage{subfig}
\usepackage[nospace,noadjust]{cite}
\usepackage{wrapfig}
\usepackage{amssymb}
\usepackage{stfloats}
\fnbelowfloat

\usepackage{etoolbox}
\makeatletter
\preto{\@verbatim}{\topsep=0pt \partopsep=0pt }
\makeatother

{\newcounter{normalfootc}
\renewcommand{\footnote}[1]{%
    \footlabel{footsaferefiwontuse\thenormalfootc}{#1}%
    \addtocounter{normalfootc}{1}%
}}

\newcolumntype{R}[1]{>{\raggedleft\let\newline\\\arraybackslash\hspace{0pt}}m{#1}}

\begin{document}

\mainmatter

\title{A Software Framework and Datasets for the Analysis of Graph Measures on RDF Graphs}
\titlerunning{Graph-based Analysis on RDF Graphs}

\author{Matth\"aus Zloch\inst{1}, Maribel Acosta\inst{2}, Daniel Hienert\inst{1}, \\Stefan Dietze\inst{1}, Stefan Conrad\inst{3}}
\institute{GESIS - Leibniz-Institute for the Social Sciences, Germany\\\email{\{firstname.lastname\}@gesis.org} \and Institute AIFB, Karlsruhe Institute of Technology, Germany\\\email{maribel.acosta@kit.edu} \and Institute for Computer Science, Heinrich-Heine University D\"usseldorf, Germany\\\email{conrad@cs.uni-duesseldorf.de}}

\authorrunning{M. Zloch, M. Acosta, D. Hienert, S. Dietze, S. Conrad}

\renewcommand*{\thefootnote}{\fnsymbol{footnote}}
\maketitle
\renewcommand*{\thefootnote}{\arabic{footnote}}

\begin{abstract}
As the availability and the inter-connectivity of RDF datasets grow, so does the necessity to understand the structure of the data. 
Understanding the topology of RDF graphs can guide and inform the development of, e.g. synthetic dataset generators, sampling methods, index structures, or query optimizers.
In this work, we propose two resources: 
(i) a software framework\footnote{\label{foot:framework-doi-link}Resource URL of the framework: \url{https://doi.org/10.5281/zenodo.2109469}} able to acquire, prepare, and perform a graph-based analysis on the topology of large RDF graphs, and 
(ii) results on a graph-based analysis of 280 datasets\footnote{\label{foot:link-doi-datasets}Resource URL of the datasets: \url{https://doi.org/10.5281/zenodo.1214433}} from the LOD Cloud with values for 28 graph measures computed with the framework.
We present a preliminary analysis based on the proposed resources and point out implications for synthetic dataset generators. Finally, we identify a set of measures, that can be used to characterize graphs in the Semantic Web.

\end{abstract}

\section{Introduction}
\label{sec:introduction}

Since its first version in 2007, the Linked Open Data Cloud (LOD Cloud) has increased by the factor of $100$, containing $1,163$ data sets in the last version of August 2017\footnote{\label{foot:lod-cloud}\url{http://lod-cloud.net/}}. 
In various knowledge domains, like Government, Life Sciences, and Natural Science, it has been a prominent example and a reference for the success of the possibility to interlink and access open datasets that are described following the Resource Description Framework (RDF).
RDF provides a graph-based data model where statements are modelled as triples. %
Furthermore, a set of RDF triples compose a directed and labelled graph, where subjects and objects can be defined as vertices while predicates correspond to edges.

Previous empirical studies on the characteristics of real-world RDF graphs have focused on general properties of the graphs~\cite{schmachtenberg2014}, or analyses on the instance or schema level of such data sets~\cite{demter2012,mihindukulasooriya2015}. Examples of statistics are dataset size, property and vocabulary usage, data types used or average length of string literals. 
In terms of the topology of RDF graphs, previous works report on network measures mainly focusing on in- and out-degree distributions, reciprocity, and path lengths~\cite{bachlechner2007,fernandez2018,flores,theoharis2008}. 
Nonetheless, the results of these studies are limited to a small fraction of the RDF datasets currently available. 

Conducting recurrent systematical analyses on a large set of RDF graph topologies is beneficial in many research areas. For instance:

\vspace{0.25em}
\noindent \textbf{Synthetic Dataset Generation.}
One goal of benchmark suites is to emulate real-world datasets and queries with characteristics from a particular domain or application-specific characteristics. 
Beyond parameters like the dataset size that is typically interpreted as the number of triples, taking into consideration reliable statistics about the network topology, basic graph and degree-based measures for instance, enables synthetic dataset generators to more appropriately emulate datasets at large-scale, contributing to solve the dataset scaling problem \cite{tay2011}.

\vspace{0.25em}
\noindent \textbf{Graph Sampling.}
At the same time, graph sampling techniques try to find a representative sample from an original dataset, with respect to different aspects. Questions that arise in this field are (1) how to obtain a (minimal) representative sample, (2) which sampling method to use, and (3) how to scale up measurements of the sample \cite{leskovec2006}. Apart from qualitative aspects, like classes, properties, instances, and used vocabularies and ontologies, also topological characteristics of the original RDF graph should be considered. To this end, primitive measures of the graphs, like the max in-, out- and average-degree of vertices, reciprocity, density, etc., may be consulted to achieve more accurate results.

\vspace{0.25em}
\noindent \textbf{Profiling and Evolution.}
Due to its distributed and dynamic nature, monitoring the development of the LOD Cloud has been a challenge for some time, documented through a range of techniques for profiling datasets \cite{benellefi2018}. Apart from the number of datasets in the LOD Cloud, the aspect of its linkage (linking into other datasets) and connectivity (linking within one dataset) is of particular interest. From the graph perspective, the creation of new links has immediate impact on the characteristics of the graph. For this reason, graph measures may help to monitor changes and the impact of changes in datasets.

To support graph-based tasks in the aforementioned areas, first, we propose an open source framework which is capable of acquiring RDF datasets, efficiently preparing and computing graph measures over large RDF graphs. 
The framework is built upon state-of-the-art third-party libraries and published under MIT license.
The proposed framework reports on network measures and graph invariants, which can be categorized in five groups:
i) basic graph measures, %
ii) degree-based measures, %
iii) centrality measures, %
iv) edge-based measures, and %
v) descriptive statistical measures.  
Second, we provide a collection of 280 datasets prepared with the framework and a report on $28$ graph-based measures per dataset about the graph topology also computed with our framework.  
In this work, we present an analysis of graph measures over the aforementioned collection. This analysis involves over $11.3$ billion RDF triples from nine knowledge domains, i.e., Cross Domain, Geography, Government, Life Sciences, Linguistics, Media, Publications, Social Networking, and User Generated. 
Finally, we conduct a correlation analysis among the studied invariants to identify a representative set of graph measures to characterize RDF datasets from a graph perspective.  
In summary, the contributions of our work are:
\vspace{-0.5em}
\begin{itemize}
    \item A framework to acquire RDF datasets and compute graph measures (\textbf{\S~\ref{sec:framework}}).
    \item Results of a graph-based analysis of $280$ RDF datasets from the LOD Cloud. For each dataset, the collection includes $28$ graph measures computed with the framework (\textbf{\S~\ref{sec:analysis}}).
    \item An analysis of graph measures on real-world RDF datasets (\textbf{\S~\ref{subsec:analysis-observations}}).  
    \item A study to identify graph measures that characterize RDF datasets (\textbf{\S~\ref{subsec:useful-measures}}).
\end{itemize}

\section{Related Work}
\label{sec:related-work}

The RDF data model imposes characteristics which are not present in other graph-based data models. %
Therefore, we distinguish between works that analyze the structure of RDF datasets in terms of RDF-specific and graph  measures.  

\vspace{0.3em} \noindent
\textbf{RDF-specific Analyses.} This category includes studies about the general structure of RDF graphs at instance, schema, and metadata levels. 
Schmachtenberg et al.~\cite{schmachtenberg2014} present the status of RDF datasets in the LOD Cloud in terms of size,  linking, vocabulary usage, and metadata.   
LODStats~\cite{demter2012} and the large-scale approach  DistLODStats~\cite{sejdiu2018} report on statistics about RDF datasets  on the web, including number of triples, RDF terms, and  properties per entity, and usage of vocabularies across datasets.   
Loupe~\cite{mihindukulasooriya2015} is an online tool that reports on the  usage of classes and properties in RDF datasets.  
Fern\'andez et al.~\cite{fernandez2018} define measures to describe the relatedness between nodes and edges using subject-object, subject-predicate, and predicate-object ratios. 
Hogan et al.~\cite{hogan2010} study the distribution of RDF terms, classes, instances, and datatypes to measure the quality of public RDF data. 
In summary,  the study of RDF-specific properties of publicly available RDF datasets have been extensively covered and is currently supported by online services and tools such as LODStats and Loupe. Therefore, in addition to these works, we focus on analyzing graph invariants in RDF datasets. 

\vspace{0.3em} \noindent
\textbf{Graph-based Analyses.} In the area of structural network analysis, it is common to study the distribution of certain graph measures in order to characterize a graph. 
RDF datasets have also been subject to these studies.
The study by Ding et al.~\cite{ding06} reveals that the power-law distribution is prevalent across graph invariants in RDF graphs obtained from 1.7 million documents. 
Also, the small-world phenomenon, known from experiments on social networks were studied within the Semantic Web~\cite{bachlechner2007}.
More recently, Fern\'andez et al.~\cite{fernandez2018} have studied the structural features of real-world RDF data.
Fern\'andez et al. also propose measures in terms of in- and -out degrees for subjects, objects, and predicates and analyze the structure of $14$ RDF graphs from different knowledge domains.   
Most of these works focus on studying different in- and out-degree distributions and are limited to a rather small collection of RDF datasets.
Moreover, the work by Flores et al.~\cite{flores} analyze further relevant graph invariants in RDF graphs including $h-$index and reciprocity. The work by Flores et al. applied graph-based metrics on synthetic RDF datasets. 
Complementary to these works, we present an  study on $280$ RDF datasets from the LOD Cloud and analyze their structure based on the average degree, $h$-index, and  powerlaw exponent.   

\section{A Framework for Graph-based Analysis on RDF Data}
\label{sec:framework}
This section introduces the first resource published with this paper: the software framework. 
The main purpose of the framework is to prepare and perform a graph-based analysis on the graph topology of RDF datasets. 
One of the main challenges of the framework is to scale up to large graphs and to a high number of datasets, i.e., to compute graph metrics efficiently over current RDF graphs (hundreds of millions of edges) and in parallel with many datasets at once. 
The necessary steps to overcome these challenges are described in the following. 

 \vspace{-0.25cm}
 \subsection{Functionality}
 \label{subsec:framework-functionality}
\begin{figure}[t!]
  \centering
  \includegraphics[width=12cm]{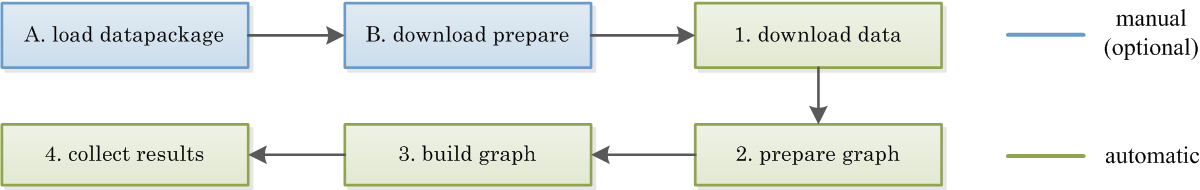}
  \caption{Illustration of the semi-automatic process pipeline. Steps 1-4 include download, data and graph preparation, and graph analysis. Steps A and B are manual and optional preparation steps, described in Section \ref{subsec:analysis-data-aquisition}.}
  \label{fig:process-pipeline}
  \vspace{-0.25cm}
 \end{figure}

 The framework relies on the following methodology to systematically acquire and analyze RDF datasets. Figure~\ref{fig:process-pipeline} depicts the main steps of our processing pipeline of the framework. In the following, we describe steps 1-4 from Figure~\ref{fig:process-pipeline}. 
 
 \subsubsection{Data Acquisition}
 \label{subsec:framework-data-acquisition}
 The framework acquires RDF data dumps available online. Online availability is not mandatory to perform  the analysis, as the pipeline runs with data dumps available offline. For convenience reasons, when operating on many datasets, one may load an initial list of datasets together with their names, available formats, and URLs into a local database (see Section \ref{subsec:analysis-data-aquisition}). One can find configuration details and database init-scripts in the source code repository\footref{foot:framework-doi-link}. Once acquired, the framework is capable of dealing with the following artifacts:
 
 \begin{itemize}
  \item Packed data dumps. Various formats are supported, including \texttt{bz2}, \texttt{7zip}, \texttt{tar.gz}, etc. This is achieved by utilizing the unix-tool \texttt{dtrx}.
  \item Archives, which contain a hierarchy of files and folders, will get scanned for files containing RDF data. Other files will be ignored, e.g. \texttt{xls},  \texttt{txt}, etc.
  \item Any files with a different serialization than N-Triples are transformed (if necessary). 
  The list of supported formats\footnote{\label{foot:rdf-formats}\url{https://www.w3.org/standards/techs/rdf#w3c_all}} is currently limited to the most common ones for RDF data, which are N-Triples, RDF/XML, Turtle, N-Quads, and Notation3. This is achieved by utilizing \textit{rapper}\footnote{\label{foot:rapper}raptor2-util library, \url{http://librdf.org/raptor/rapper.html}}.
 \end{itemize}

 \vspace{-2mm}
 \vspace{-0.4cm}
 \subsubsection{Preparation of the Graph Structure}
 \label{subsec:framework-graph-prepare}
 In order to deal with large RDF graphs, our aim is to create a as much automated and reliable processing pipeline as possible that focuses on performance. 
 The graph structure is created from an edgelist, which is the result of this preparation step. One line in the edgelist constitutes one edge in the graph, which is a relation between a pair of vertices, the subject \texttt{s} and object \texttt{o} of an RDF triple. The line contains the predicate \texttt{p} of an RDF triple in addition, so that it is stored as an attribute of the edge. This attribute can be accessed during graph analysis and processing.
 To ease the creation of this edgelist with edge attributes, we utilized the N-Triples format, thus, a triple \texttt{s p o} becomes \texttt{s o p} in the edgelist. By this means, the framework is able to prepare several datasets in parallel.
 
 In order to reduce the usage of hard-disk space and also main memory during the creation process of the graph structure, we make use of an efficient state-of-the-art non-cryptographic hashing function\footnote{\label{foot:xxhash-link}xxhash, \url{https://github.com/Cyan4973/xxHash}} to encode actual values of the RDF triples. For example, the RDF triple

 \vspace{0.5em}
 {\small
 \begin{verbatim}
    <http://data.linkedopendata.it/musei/resource/Roma> 
         <http://www.w3.org/2000/01/rdf-schema#label> "Roma" .
 \end{verbatim}}
 \vspace{-0.25cm}

is turned into the hashed edgelist representation

 \vspace{0.5em}
 {\small \hspace{1em}\texttt{43f2f4f2e41ae099 c9643559faeed68e 02325f53aeba2f02}}
 \vspace{0.5em}
 
 Besides the fact that this hashing strategy can reduce space by the factor of up to 12, compared to simple integer representation it has the advantage that it facilitates the comparison between edgelists of different RDF datasets. One could examine which resource URIs are the most frequently used across all datasets. The framework provides a script to de-reference hashes, in order to find a resource URI for the vertex with maximum degree, for instance.

 \vspace{-0.4cm}
 \subsubsection{Graph Creation}
 \label{subsec:framework-graph-creation}
 As graph analysis library we used \textit{graph-tool}\footnote{\label{foot:graph-tool-link}graph-tool, \url{https://graph-tool.skewed.de/}}, an efficient library for statistical analysis of graphs. In \textit{graph-tool}, core data structures and algorithms are implemented in C$^{++}$/C, while the library itself can be used with Python.  \textit{graph-tool} comes with a lot of pre-defined implementations for graph analysis, e.g.,  degree distributions or more advanced implementations on graphs like PageRank or clustering coefficient. Further, some values may be stored as attributes of vertices or edges in the graph structure. 
 
 The library's internal graph-structure may be serialized as a  compressed binary object for future re-use. It can be reloaded by \textit{graph-tool} with much higher performance than the original edgelist.
 Our framework instantiates the graph from the prepared edgelist or binary representation and operates on the graph object provided by the \textit{graph-tool} library. 
 As with dataset preparation, the framework can handle multiple computations of graph measures in parallel.
 
 \subsection{Graph Measures}
 \label{subsec:framework-measures}

 In this section, we present statistical measures that are computed in the framework grouped into five dimensions: basic graph measures, degree-based measures, centrality measures, edge-based measures, and descriptive statistical measures. The computation of some metrics are carried out with \textit{graph-tool} (e.g., PageRank), and others are computed by our framework (e.g., degree of centralization).
 
 In the following, we introduce the graph notation used  throughout the paper. 
 A graph $G$ is a pair of finite sets ($V$, $E$), with $V$ denoting the set of all vertices (RDF subject and object resources). $E$ is a multiset of (labeled) edges in the graph $G$, since in RDF a pair of subject and object resources may be described with more than one predicate. E.g. in the graph {\small \texttt{\{ s p1 o. s p2 o \}}}, $E$ has two pairs of vertices, i.e. $E=\{(s,o)_1,(s,o)_2\: |\: s,o \in V\}$. 
 RDF predicates are considered as additional edge labels, which also may occur as individual vertices in the same graph $G$. 
 Newman~\cite{newman2010} presents a more detailed introduction to networks and structural network analysis. 
 
 \vspace{-0.4cm}
 \subsubsection{Basic Graph Measures}
 \label{subsubsec:measures-basic-graph}
 We report on the total number of vertices $|V| = n$ and the number of edges $|E| = m$ for a graph. Some works in the literature refer to these values as size and volume, respectively. The number of vertices and edges usually varies drastically across knowledge domains.
 
 By its nature, RDF graphs contain a fraction of edges that share the same pair of source and target vertices (as in the example above). In our work, $m_p$ represents the number of parallel edges, i.e., $m_p = |\{ e\in E\: |\: count(e,E) > 1 \}|$, with $count(e,E)$ being a function that returns the number of times $e$ is contained in $E$. %
 Based on this measure, we also compute the total number of edges without counting parallel edges, denoted $m_u$. It is computed by subtracting $m_p$ from the total number of edges $m$, i.e., $m_u = m - m_p$.
 
 \vspace{-0.4cm}
 \subsubsection{Degree-based Measures}
 \label{subsubsec:degree-based-measures}
 The degree of a vertex $v \in V$, denoted $d(v)$, corresponds to the total number of incoming and outgoing edges of $v$, i.e., $d(v) = |\{ (u,v) \in E \ or \  (v,u) \in E\: |\: u \in V \}|$. For directed graphs, as is true for RDF datasets, it is common to distinguish between in- and out-degree, i.e. $d_{in}(v) = |\{ (u,v) \in E\: |\: u \in V \}|$ and $d_{out}(v) = |\{ (v,u) \in E\: |\: u \in V \}|$, respectively.
 In social network analyses, vertices with a high out-degree are said to be ``influential", whereas vertices with a high in-degree are called ``prestigious". 
 To identify these vertices in RDF graphs, we compute the maximum total-, in-, and out-degree of the graph's vertices, i.e., $d_{max} = max\ d(v)$, $d_{max,in} = max\ d_{in}(v)$, $d_{max,out} = max\ d_{out}(v)$, $\forall v \in V$ respectively. 
 In addition, we compute the graph's average total-, in-, and out-degree denoted $z$, $z_{in}$, and $z_{out}$, respectively. 
 These measures can be important in research on RDF data management, for instance, where the (average) degree of a vertex (database table record) has significant impact on query evaluation, since queries on dense graphs can be more costly in terms of execution time to evaluate \cite{qiao2015}.

 Another degree-based measure supported in the framework is $h-$index, known from citation networks~\cite{hirsch2005}. It is an indicator for the importance of a vertex, similar to a centrality measure (see Section \ref{subsubsec:centrality-measures}). A value of $h$ means that for the number of $h$ vertices the degree of these vertices is greater or equal to $h$. A high value of a graph's $h-$index could be an indicator for a ``dense" graph and that its vertices are more ``prestigious". We compute this network measure for the directed graph (using only the in-degree of vertices) denoted as $h_d$ and the undirected graph (using in- and out-degree of vertices) denoted as $h_u$.
 
 \vspace{-0.4cm}
 \subsubsection{Centrality Measures}
 \label{subsubsec:centrality-measures}
 In social network analyses, the concept of \textit{point centrality} is used to express the importance of nodes in a network. There are many interpretations for the term ``importance" and so are measures for centrality \cite{newman2010}.
 Comparing centrality measures with fill $p$ shows that the higher the density of the graph the higher centrality measures it has for the vertices.
 Point centrality uses the degree of a vertex, $d(v)$. To indicate that it is a centrality measure, the literature sometimes normalizes this values by the total number of all vertices. We compute the maximum value of this measure, denoted as $C_{D,max}~=~d_{max}$.
 Another centrality measure computed is PageRank~\cite{page1999}. For each RDF graph, we identified the vertex with the highest PageRank values, denoted as $PR_{max}$. 
     
 Besides the point centrality, there is also the measure of \textit{graph centralization}~\cite{freeman1979}, which is known from social network analysis. This measure may also be seen as an indicator for the type of the graph, in that it expresses the degree of inequality and concentration of vertices as can be found in a perfect star-shaped graphs, that is at most centralized and unequal with regard to its degree distribution. The centralization of a graph regarding the degree is defined as:
 
 \begin{equation}
     C_D = \frac{\sum_{}^{v\in V} (d_{max} - d(v))}{(|V|-1)*(|V|-2)}
 \end{equation}
 
 \noindent
 where $C_D$ denotes the graph centralization measure using degree~\cite{freeman1979}. In contrast to social networks, RDF graphs usually contain many parallel edges between vertices (see next subsection). Thus, for this measure to make sense, we used the number of unique edges in the graph, $m_u$.
 
 \vspace{-0.4cm}
 \subsubsection{Edge-based Measures}
 \label{subsubsec:edges-based-measures}
 We compute the ``density" or ``connectance" of a graph, called $fill$ denoted as $p$. It also can be interpreted as the probability that an edge is present between two randomly chosen vertices. 
The density is computed as the ratio of all edges to the total number of all possible edges. We use the formula for a directed graph with possible loops, accordance to the definition of RDF graphs, using $m$ and $m_u$, i.e. $p = m/n^2$ and $p_u = m_u/n^2$. 
 
 Further, we analyze the fraction of bidirectional connections between vertices in the graph, thus pairs of vertices forward-connected by some edge, which are also backward-connected by some other edge. The value of \textit{reciprocity}, denoted as $y$, is expressed as percentage, i.e. $y = m_{bi}/m$, with $m_{bi} = |\{ (u,v) \in E\: |\: \exists (v,u) \in E \}|$. A high value means there are many connections between vertices which are bidirectional. This value is expected to be high in citation or social networks.
 
 Another important group of measures that is described by the graph topology is related to paths. A path is a set of edges one can follow along between two vertices. As there can be more than one path, the \textit{diameter} is defined as the longest shortest path between two vertices of the network~\cite{newman2010}, denoted as $\delta$. This is a valuable measure when storing an RDF dataset in a relational database, as this measure affects join cardinality estimations depending on the type of schema implementation for the graph set. 
 The diameter is usually a very time consuming measure to compute, since all possible paths have to be computed. Thus we used the \texttt{pseudo\_diameter} algorithm\footnote{\label{foot:graph-tool-pseudo-diameter}\url{https://graph-tool.skewed.de/static/doc/topology.html#graph_tool.topology.pseudo_diameter}} to estimate the value for our datasets.

 \vspace{-0.4cm}
 \subsubsection{Descriptive Statistical Measures}
 \label{subsubsec:descriptive-statistical-measures}
 Descriptive statistical measures are important to describe distributions of some set of values, in our scenario, values for graph measures. In statistics, it is common to compute the $variance\ \sigma^2$ and $standard\ deviation\ \sigma$ in order to express the degree of dispersion of a distribution. We do this for the in- and out-degree distributions in the graphs, denoted by $\sigma^2_{in}$, $\sigma^2_{out}$ and $\sigma_{in}$, $\sigma_{out}$, respectively. Furthermore, the~\textit{coefficient of variation cv} is consulted to have a comparable measure for distributions with different mean values. $cv_{in}$ and $cv_{out}$ are obtained by dividing the corresponding standard deviation $\sigma_{in}$ and $\sigma_{out}$ by the mean $z_{in}$, and $z_{out}$, respectively, times 100. $cv$ can also be utilized to analyze the type of distribution with regard to a set of values. For example, a low value of $cv_{out}$ means constant influence of vertices in the graph (homogeneous group), whereas a high value of $cv_{in}$ means high prominence of some vertices in the graph (heterogeneous group). 
 
 Further, the type of \textit{degree} distribution is an often considered measure of graphs. Some domains and datasets report on degree distributions that follow a power-law function, which means that the number of vertices with degree $k$ behaves proportionally to the power of $k^{-\alpha}$, for some $\alpha \in \mathbb{R}$. Such networks are called scale-free. The literature has found that values in the range of 2 $< \alpha <$ 3 are typical in many real-world networks~\cite{newman2010}. The scale-free behaviour also applies to some datasets and measures of RDF datasets~\cite{fernandez2018,ding06}. However, to reason about whether a distribution follows a power-law can be technically challenging~\cite{alstott2014}, and computing the exponent $\alpha$, that falls into a certain range of values, is not sufficient. 
 We compute the exponent for the total- and in-degree distributions~\cite{alstott2014}, denoted as $\alpha$ and $\alpha_{in}$, respectively. In addition, to support the analysis of  power-law distributions, the framework produces plots for both distributions. A power-law distribution is described as a line in a log-log plot. 
 
 Determining the function that fits the distribution may be of high value for algorithms, in order to estimate the selectivity of vertices and attributes in graphs. The structure and size of datasets created by synthetic datasets, for instance, can be controlled with these measures. Also, a clear power-law distribution allows for high compression rates of RDF datasets~\cite{fernandez2018}.
 
 \vspace{-0.25cm}
 \subsection{Availability, Sustainability and Maintenance}
 \label{subsec:framework-maintenance}
 The software framework is published under MIT license on GitHub\footref{foot:framework-doi-link}. The repository contains all code and a comprehensive documentation to install, prepare an RDF dataset, and run the analysis. The main part of the code  implements most of the measures as a list of python functions that is extendable. Future features and bugfixes will be published under a minor or bugfix release, v0.x.x, respectively.
 The source code is frequently maintained and debugged, since it is actively used in other research projects at our institute (see Section \ref{sec:conclusion}). It is citable via a registered DOI obtained from Zenodo. Both web services, GitHub and Zenodo, provide search interfaces, which makes the code also be web findable.
 
\section{RDF Datasets for the Analysis of Graph Measures}
\label{sec:analysis}
We conducted a systematic graph-based analysis with a large group of datasets which were part of the last LOD Cloud 2017\footnote{\label{foot:link-lod-cloud-2017}\url{http://lod-cloud.net/versions/2017-08-22/datasets_22-08-2017.tsv}.}, as a case study for the framework introduced in the previous Section \ref{sec:framework}. 
The results of the graph-based analysis with 28 graph-based measures, is the second resource\footref{foot:link-doi-datasets} published with this paper.
To facilitate browsing of the data we provide a website\footnote{\label{foot:link-project-website}\url{http://data.gesis.org/lodcc/2017-08}}. It contains all 280 datasets that were analyzed, grouped by topics (as in the LOD Cloud) together with links (a) to the original metadata obtained from DataHub, and (b) a downloadable version of the serialized graph-structure used for the analysis.
This section describes the data acquisition process (cf. sections \ref{subsec:analysis-data-aquisition} and \ref{subsec:analysis-execution-environment}), and how the datasets and the results of the analysis can be accessed (cf. Section \ref{subsec:analysis-maintenance}). 

 \vspace{-0.25cm}
 \subsection{Data Acquisition}
 \label{subsec:analysis-data-aquisition}
 Table \ref{table:largest-datasets} summarizes the number of processed datasets and their sizes. From the total number of 1,163 potentially available datasets in last LOD Cloud 2017, a number of 280 datasets were in fact analyzed. This was mainly due to these reasons: (i) RDF media types statements that were actually correct for the datasets, and (ii) the availability of data dumps provided by the services. To not stress SPARQL endpoints to transfer large amounts of data, in this experiment, only datasets that provide downloadable dumps were considered.

 \begin{table}[t!]
  \vspace{-0.4cm}
  \caption{Processed datasets. Number of datasets, average and maximum number of vertices ($n$) and edges ($m$) in RDF graphs per knowledge domain. }\vspace{0.5em}
  \scriptsize
  \centering
  \setlength{\tabcolsep}{0.6em}
  {\renewcommand{\arraystretch}{1.1}
  \begin{tabular}{ | l | r | r | r | r | R{1.8cm} | r |}
    \hline
     \textbf{Domain} & \multicolumn{2}{l|}{\textbf{Maximum}} & \multicolumn{2}{l|}{\textbf{Average}} & \textbf{\# datasets} \\
     \cline{2-5}
     & $\boldsymbol{n}$ & $\boldsymbol{m}$ & $\boldsymbol{n}$ & $\boldsymbol{m}$ & \\
    \hline
     Cross Domain & 614,448,283 & 2,656,226,986 & 57,827,358 & 218,930,066 & 15 \\
    \hline
     Geography & 47,541,174 & 340,880,391 & 9,763,721 & 61,049,429 & 11 \\
    \hline
     Government & 131,634,287 & 1,489,689,235 & 7,491,531 & 71,263,878 & 37 \\
    \hline
     Life Sciences & 356,837,444 & 722,889,087 & 25,550,646 & 85,262,882 & 32 \\
    \hline
     Linguistics & 120,683,397 & 291,314,466 & 1,260,455 & 3,347,268 & 122 \\
    \hline
     Media & 48,318,259 & 161,749,815 & 9,504,622 & 31,100,859 & 6 \\
    \hline
     Publications & 218,757,266 & 720,668,819 & 9,036,204 & 28,017,502 & 50 \\
    \hline
     Social Networking & 331,647 & 1,600,499 & 237,003 & 1,062,986 & 3 \\
    \hline
     User Generated & 2,961,628 & 4,932,352 & 967,798 & 1,992,069 & 4 \\
    \hline
  \end{tabular} \vspace{-0.4cm} }
 \label{table:largest-datasets}
 \end{table}
 
 To dereference RDF datasets we relied on the metadata (so called data-package) available at DataHub, which specifies URLs and media types for the corresponding data provider of one dataset\footnote{\label{foot:datahub-datapackage}Example: \url{https://old.datahub.io/dataset/<dataset-name>/datapackage.json}}. We collected the datapackage metadata for all datasets (step A in Figure~\ref{fig:process-pipeline}) and manually mapped the obtained media types from the datapackage to their corresponding official media type statements that are given in the specifications. For instance, \texttt{rdf}, \texttt{xml\_rdf} or \texttt{rdf\_xml} was mapped to \texttt{application/rdf+xml} and similar. Other media type statements like~\texttt{html\_json\_ld\_ttl\_rdf\_xml} or~\texttt{rdf\_xml\_turtle\_html} were ignored, since they are ambiguous. This way, we obtained the URLs of 890 RDF datasets (step B in Figure \ref{fig:process-pipeline}). After that, we checked whether the dumps are available by performing \texttt{HTTP HEAD} requests on the URLs. At the time of the experiment, this returned 486 potential  RDF dataset dumps to download. %
 For the other not available URLs we verified the status of those datasets with {\small\url{http://stats.lod2.eu}}. After these manual preparation steps the data dumps could be downloaded with the framework (step 1 in Figure \ref{fig:process-pipeline}).
 
 The framework needs to transform all formats into N-Triples (cf. Section \ref{subsec:framework-graph-prepare}). From here, the number of prepared datasets for the analysis further reduced to 280. The reasons were: (1) corrupt downloads, (2) wrong file media type statements, and (3) syntax errors or other formats than these what were expected during the transformation process. This number seems low compared to the total number of available datasets in the LOD cloud, although it sounds reasonable compared to a recent study on the LOD Cloud 2014 \cite{debattista2018}. 

 \vspace{-0.25cm}
 \subsection{Execution Environment}
 \label{subsec:analysis-execution-environment}
 Operating system, database installation, dataset, and client software reside all on one server during analysis. The analysis was made on a rack server Dell PowerBridge R720, having two Intel(R) Xeon(R) E5-2600 processors with 16 cores each, 192GB of main memory, and a 5TB main storage. The operating system was Linux, Debian 7.11, kernel version 3.2.0.5. 
 
 The framework was configured to download and prepare the RDF data dumps in a parallel manner, limited to 28 concurrent processes, since transformation processes require some hard-disk IO. Around 2TB of hard-disk space was required to finish the preparation. The analysis on the graphs require more main memory, thus it was conducted only with 12 concurrent processes. As serialized binary objects all 280 datasets required around 38GB. Table \ref{table:analysis-times} depicts examples of times for dataset preparation and analysis in our environment.
 
 \begin{table}[t!]
   \vspace{-0.4cm}
   
   \caption{Runtime for the different stages in our graph-based analysis for selected datasets. All files needed to be transformed from RDF/XML into N-Triples. \\ \tiny{* Compressed archive with multiple RDF files.}} \vspace{0.5em}
   \centering
   \scriptsize
   \setlength{\tabcolsep}{0.5em}
   {\renewcommand{\arraystretch}{1.1}
   \begin{tabular}{ | l | r | r | r | r | }
     \hline
     \textbf{Dataset Name} & $\boldsymbol{m}$ \textbf{Edges} & $\boldsymbol{t_1}$\textbf{ Preparation} & $\boldsymbol{t_2}$ \textbf{Graph Creation} & $\boldsymbol{t_3}$ \textbf{Graph Analysis} \\
     \hline
     \textit{colinda} & 100,000 & 2.26s & 0.67s & 3.62s \\
     \hline
     \textit{organic-edunet} & 1,200,000 & 25.81s & 8.62s & 16.95s \\
     \hline
     \textit{uis-linked-data} & *10,300,000 & 203.05s & 61.01s & 26.13s \\
     \hline
   \end{tabular} }
   \label{table:analysis-times}
   \vspace{-0.4cm}
 \end{table}

 \subsection{Availability, Sustainability and Maintenance}
 \label{subsec:analysis-maintenance}
 The results of the analysis of 280 datasets with 28 graph-based measures and degree distribution plots per dataset can be examined and downloaded via the registered DOI\footref{foot:link-doi-datasets}. 
 The aforementioned website\footref{foot:link-doi-datasets} is automatically generated from the results. It contains all 280 datasets that were analyzed, grouped by topic domains (as in the LOD Cloud) together with links (a) to the original metadata obtained from datahub and (b) a downloadable version of the serialized graph-structure used by the time of analysis (as described in Section \ref{subsec:framework-functionality}). 

 As an infrastructure institute for the Social Sciences, we will regularly load data from the LOD Cloud and (re-)calculate the measures for the obtained datasets. This is part of a linking strategy, where linking candidates for our datasets shall be identified\footnote{\label{foot:link-gesis-research-data}\url{https://search.gesis.org/research_data}}.
 Datasets and results of future analyses will be made available to the community for further research. 

\section{Preliminary Analysis and Discussion}
\label{sec:results}

This section presents some results and observations about RDF graph topologies in the LOD Cloud, obtained from analyzing 280 datasets with the framework, as described in the previous Section \ref{sec:analysis}. 
The interested reader is encouraged to look-up single values in the measures section of one dataset on the website of the project\footref{foot:link-project-website}. %
In the following, we present our main observations on basic graph measures, degree-based measures, and degree distribution statistics.

 \vspace{-0.25cm}
 \subsection{Observations about Graph Topologies in the LOD Cloud}
 \label{subsec:analysis-observations}
 
  \begin{figure}[t!] %
  \centering
  \includegraphics[width=1.00\textwidth]{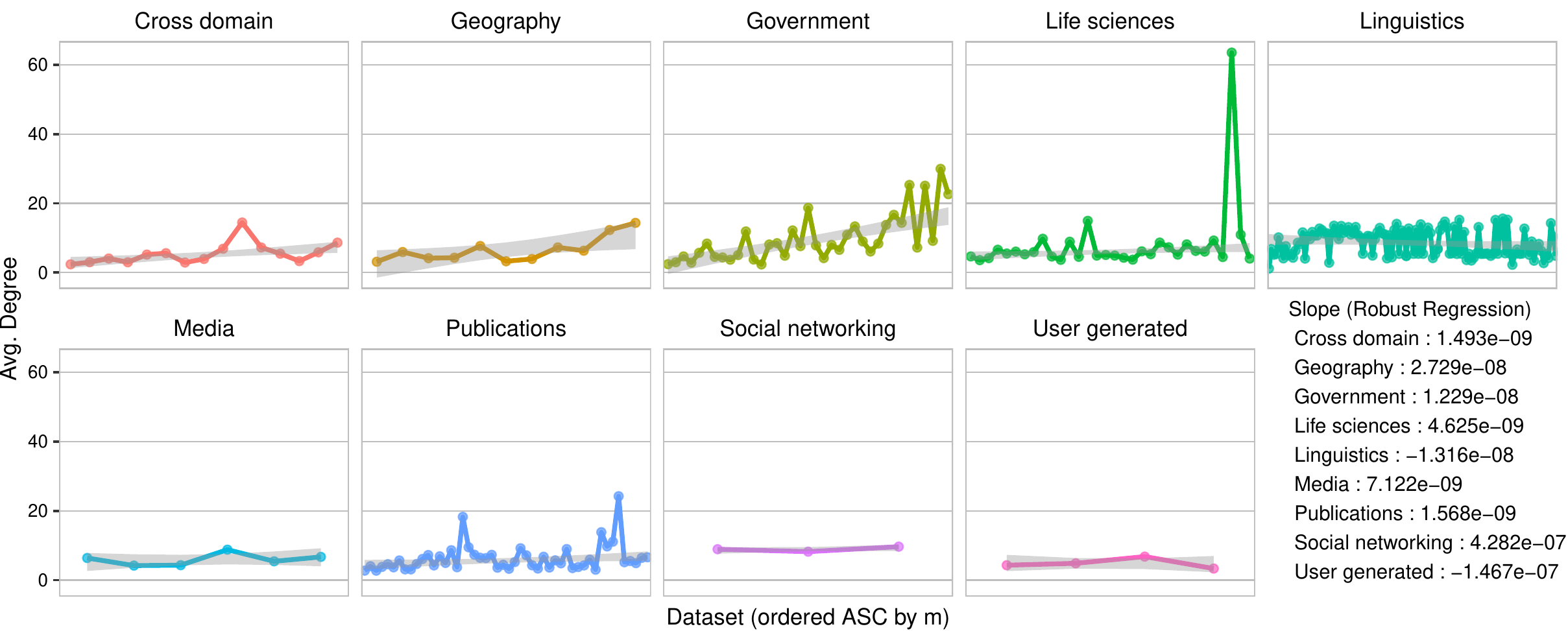} \vspace{-0.4cm}
  \caption{Average degree $z$. The x-axis is ordered by the number of edges $m$. The slope of trend lines is computed by robust regression using \textit{M-estimation}.} 
  \vspace{-0.25cm}
  \label{fig:measure-avg-degree} 
 \end{figure}
 
 \subsubsection{Basic Graph Measures}
 Figure \ref{fig:measure-avg-degree} shows the average degree of all analyzed datasets. Among all domains but Geography and Government, it seems that the average degree is not affected by the volume of the graph (number of edges). 
 Datasets in the Geography and Government domains report an increasing linear relationship with respect to the volume. Some outliers with high values can be observed across all domains, especially in Geography, Life Sciences, and Publications. 
 The highest value over all datasets can be found in the Life Sciences domain, with $63.50$ edges per vertex on average (\textit{bio2rdf-irefindex}). Over all observed domains and datasets, the value is $7.9$ on average (with a standard deviation of $1.71$).
Datasets in Cross Domain have the lowest value of $5.46$ (User Generated domain has even $4.81$, but only few datasets could be examined).

 \vspace{-0.4cm}
 \subsubsection{Degree-based Measures}
 \label{subsubsec:discussion-degree-based-measures}
\begin{figure}[t!]
  \centering
  \includegraphics[width=1.00\textwidth]{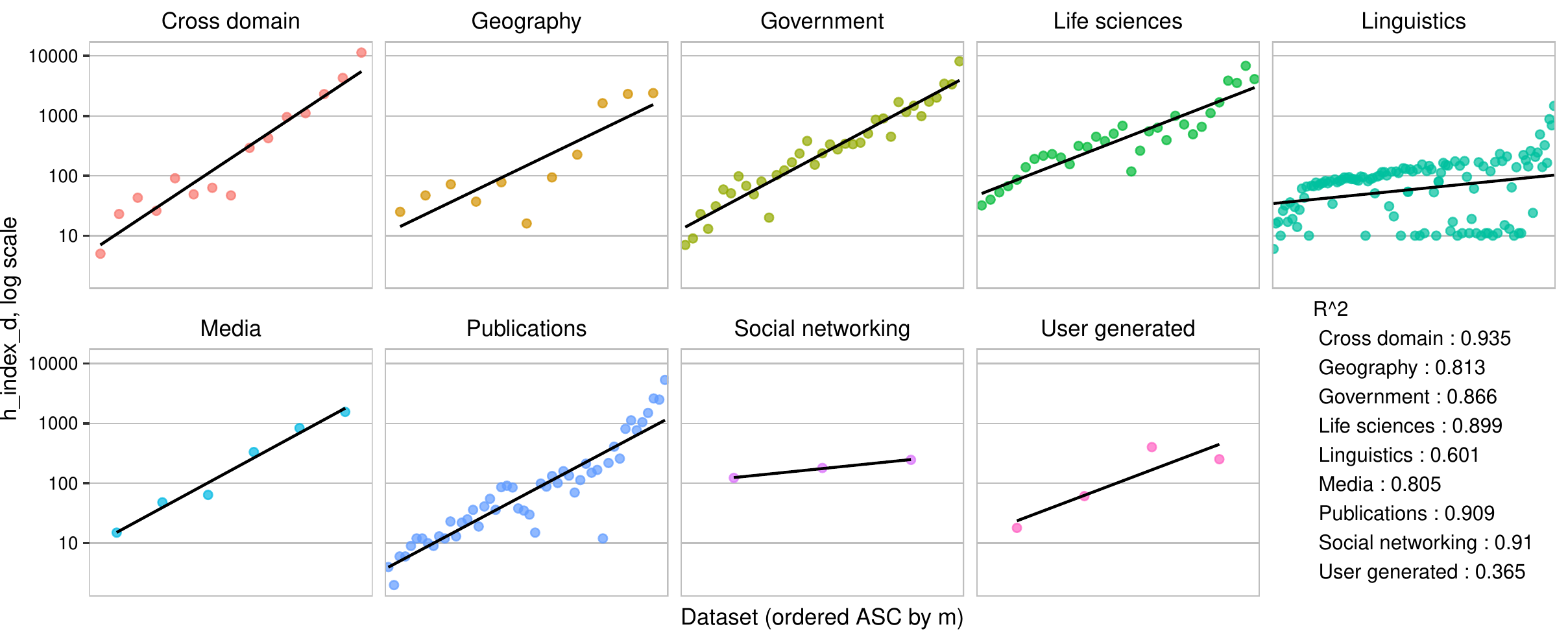} %
  \caption{$h$-index. The x-axis (log scale) is ordered by the number of edges $m$. Each plot has the same range for the x-axis. $R^2$ measures how well the regression fits. The closer to 1 the better the prediction.}
  \label{fig:measure-h-index}
  \vspace{-0.25cm}
 \end{figure}

 Figure \ref{fig:measure-h-index} shows the results on $h$-index. We would like to address some (a) domain-specific and (b) dataset-specific observations. 
 
 Regarding (a), we can see that in general, the $h$-index grows exponentially with the size of the graph (note the log-scaled y-axis). 
 Some datasets in the Government, Life Sciences, and Publications domains have high values for $h-$index: 8,128; 6,839; 5,309, respectively. Cross Domain exhibits the highest $h-$index values on average, with \textit{dbpedia-en} having the highest value of 11,363. Repeating the definition, this means that there are 11,363 vertices in the graph with at least 11,363 or more edges, which is surprising. Compared to other domains, datasets in the Linguistics domain have a fairly low $h$-index, with 115 on average (other domains at least 3 times higher). 
 
 Regarding (b), dataset-specific phenomena can be observed in the Linguistics domain. There seems to be two groups with totally different values, obviously due to datasets with very different graph topology. 
 In this domain,  \textit{universal-dependencies-treebank} is present with 63 datasets, \textit{apertium-rdf} with 22 datasets. 
 Looking at the actual values for these groups of datasets, we can see that \textit{apertium-rdf} datasets are 6x larger in size (vertices) and 2.6x larger in volume (edges) than \textit{universal-dependencies-treebank}. The average degree in the first group is half the value of the second group (5.43 vs. 11.62). However, their size and volume seems to have no effect on the values of $h$-index. The first group of datasets have almost constant $h$-index value (lower group of dots in the figure), which is 10x smaller on average than that of datasets of \textit{universal-dependencies-treebank} (upper group of dots). This, obviously, is not a domain-specific, but rather a dataset-specific phenomenon. 
 
 \vspace{-0.4cm}
 \subsubsection{Degree Distribution Statistics}

\begin{figure}[t!]
  \centering
   \subfloat[\scriptsize panlex \newline domain: Linguistics \newline $n\ \sim\ 120m; m\ \sim\ 291m$ \newline $\alpha\ =\ 2.43; dmin\ =\ 39$]{
   \includegraphics[width=0.32\textwidth]{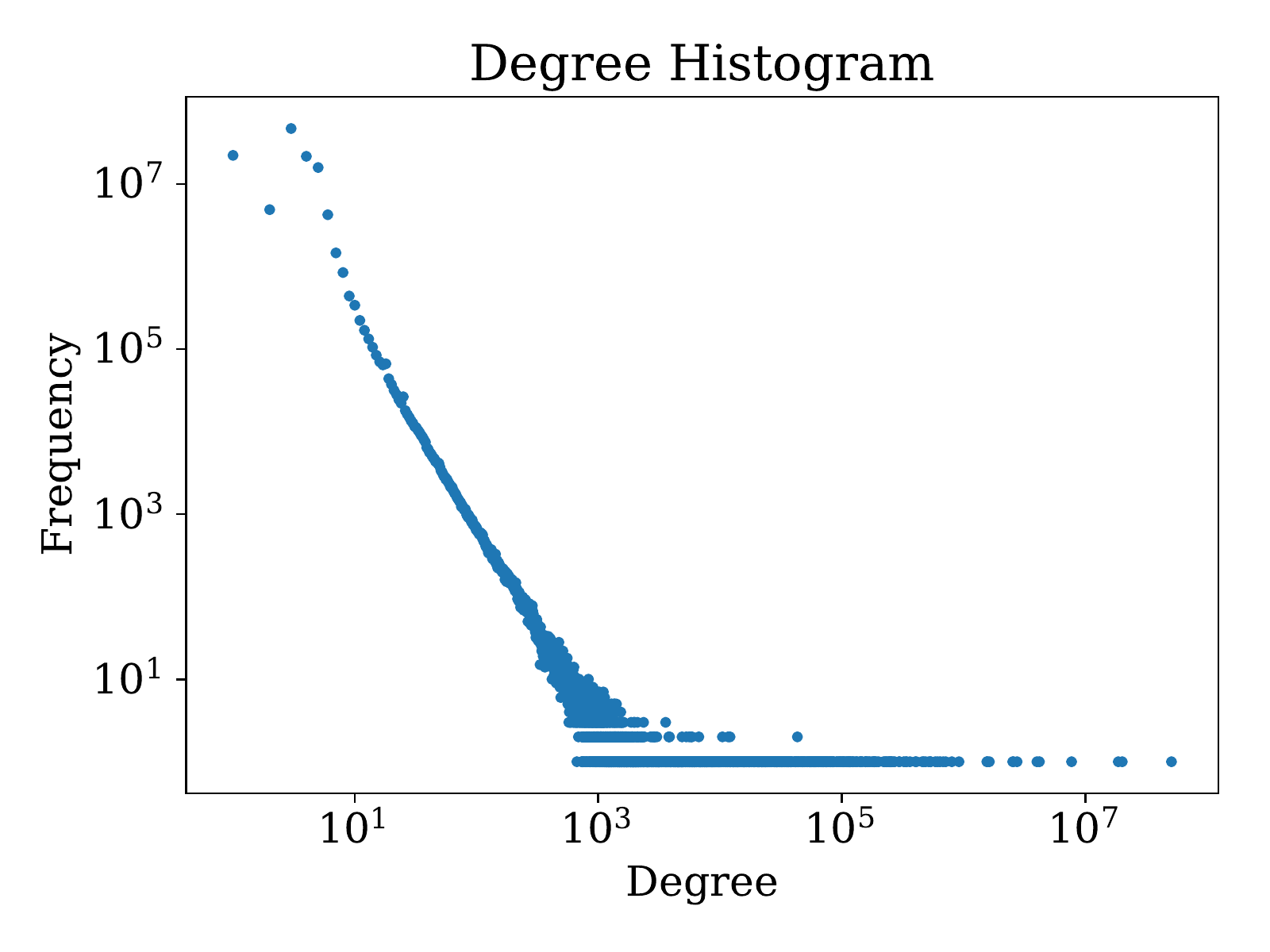}
   \label{fig:plot-panlex-distribution}}
   \subfloat[\scriptsize rkb-expl.-webconf \newline domain: Publications \newline $n\ \sim\ 57k; m\ \sim\ 205k$ \newline $\alpha\ =\ 2.98; dmin\ =\ 22$]{
   \includegraphics[width=0.32\textwidth]{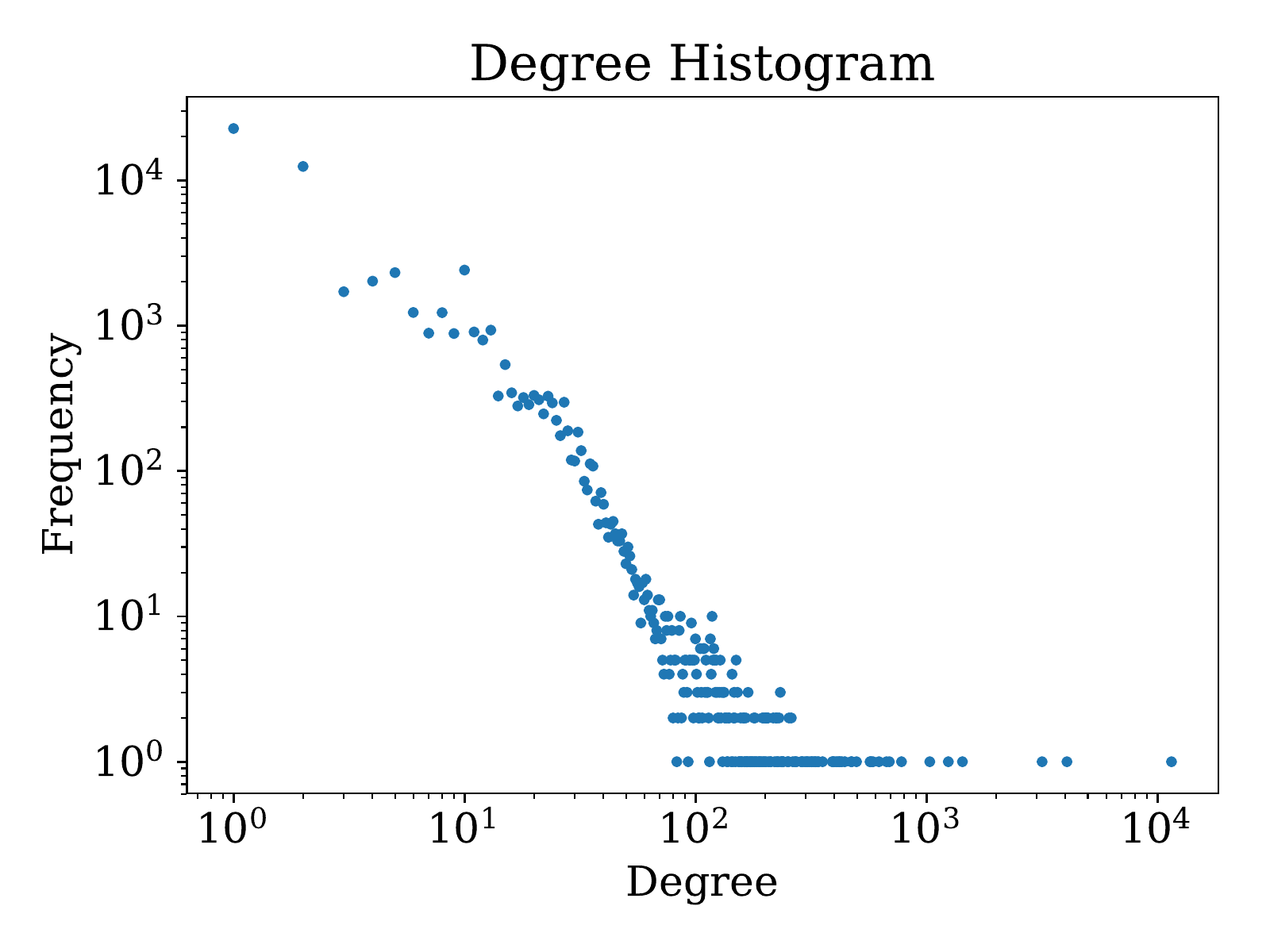}
   \label{fig:plot-webconf-distribution}}
   \subfloat[\scriptsize core \newline domain: Publications \newline $n\ \sim\ 1.56m; m\ \sim\ 3.3m$ \newline $\alpha\ =\ 24.91; dmin\ =\ 4$]{
   \includegraphics[width=0.32\textwidth]{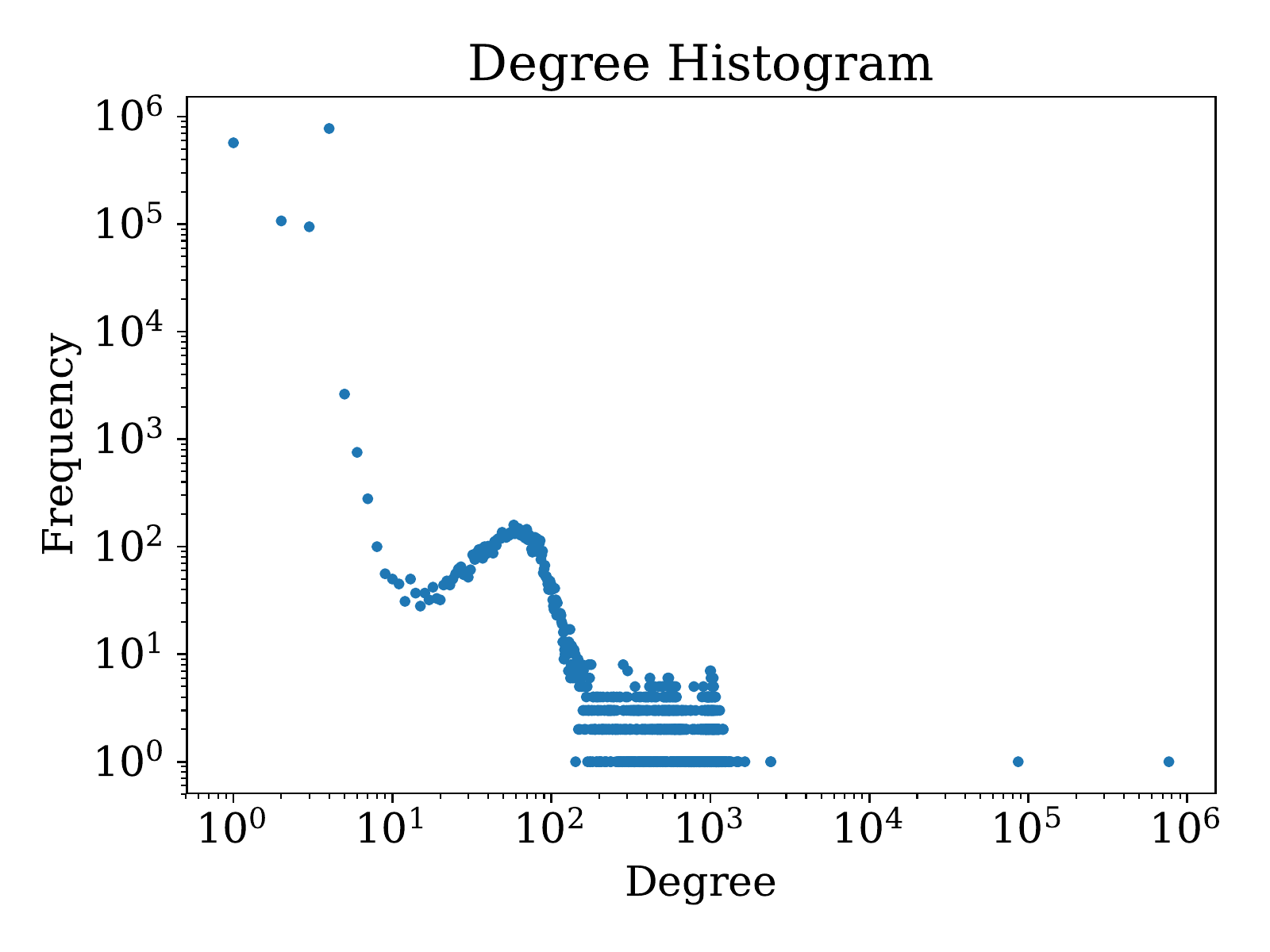}
   \label{fig:plot-core-distribution}}
  
  \caption{Exemplary plots created by the framework for datasets of different sizes. Please note the double-logarithmic axes. $dmin$ is the position on the x-axis from which the data may fit a power-law function.}
  \label{fig:plots-degree-distributions}
  \vspace{-0.4cm}
 \end{figure}

 Researchers have found scale-free networks and graphs in many datasets~\cite{newman2010,ding06}, with a power-law exponent value of 2 $< \alpha <$ 3. We can confirm that for many of the analyzed datasets. As described in Section \ref{subsubsec:descriptive-statistical-measures}, it is generally not sufficient to decide whether a distribution fits a power-law function just by determining the value of $\alpha$. Exemplary plots created by the framework for graphs of different sizes are presented in Figures \ref{fig:plot-panlex-distribution} and \ref{fig:plot-webconf-distribution}. These graphs reveal a scale-free behaviour with 2 $< \alpha <$ 3 for their degree distribution. Figure \ref{fig:plot-core-distribution} is an example for a degree-distribution not following a powerlaw function. For a detailed study on the distributions please find plots for all analyzed datasets on the website of our project\footref{foot:link-project-website}. 
 
 Looking at the actual data for all datasets, we could observe that, in general, values for exponent $\alpha$ and for $d_{min}$ vary a lot across domains. Furthermore, many datasets exhibit a scale-free behaviour on the total-degree distributions, but not on the in-degree, and vice-versa. It is hard to tell if a scale-free behaviour is a characteristic for a certain domain. We came to the conclusion that this is a dataset-specific phenomenon. However, the Publications domain has the highest share of datasets with 2 $< \alpha <$ 3 for total- and in-degree distributions, i.e., 62\% and 74\%, respectively.

 \subsection{Effective Measures for RDF Graph Analysis}
 \label{subsec:useful-measures}
 Regarding the aforementioned use case of synthetic dataset generation, one goal of benchmark suites is to emulate real-world datasets with characteristics from a particular domain. Typical usages of benchmark suites is the study of runtime performance of common (domain-specific) queries at large scale. Some of them have been criticized to not necessarily generate meaningful results, due to the fact that datasets and queries are artificial with little relation to real datasets~\cite{duan2011}. 

 \begin{wrapfigure}[14]{R}{0.42\textwidth}
     \vspace{-8mm}
     \centering
     \includegraphics[width=0.40\textwidth]{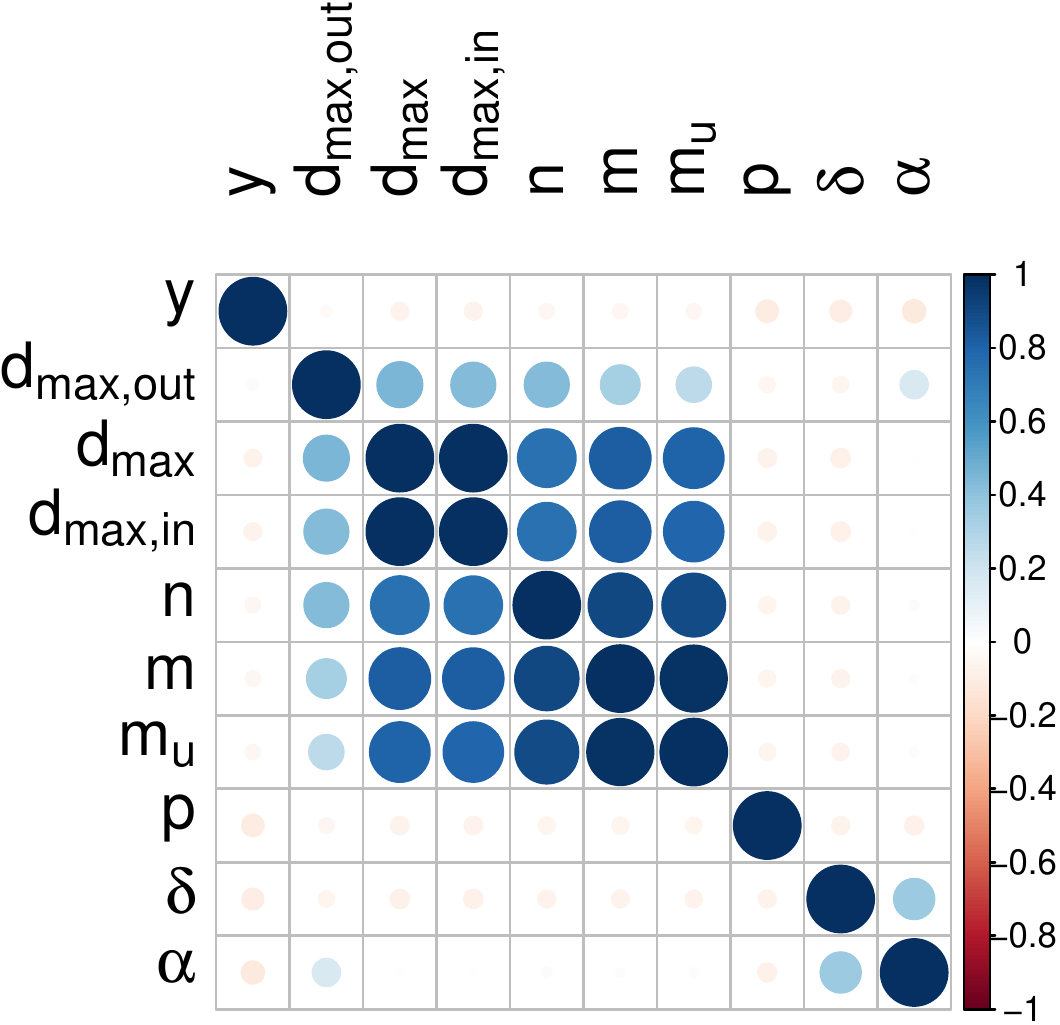}
     \caption{Measure correlation}
     \label{fig:answerdist}
 \end{wrapfigure}
 \noindent Recent works are proposing a paradigm shift from domain-specific benchmarks, which utilize a predefined schema and domain-specific data, towards designing application-specific benchmarks \cite{tay2011,qiao2015}. We have observed such discrepancies in the Linguistics domain, for instance (cf. Section \ref{subsubsec:discussion-degree-based-measures}).
 For both approaches, the results of our framework could facilitate the development of more accurate results, by combining topological measures, like the ones that can be obtained by the framework presented in this paper, with measures that describe statistics of vocabulary usage, for instance.

 One may come to the question, which measures are essential for graph characterization. We noticed that many measures rely on the degree of a vertex. A Pearson correlation test on the results of the analysis of datasets from Section \ref{sec:analysis} shows that $n$, $m$, $m_u$, and $m_p$, correlate strongly to both $h$-index measures and to the standard descriptive statistical measure. 
 The degree of centralization and degree centrality correlates with $d_{max}$, $d_{max,in}$, $d_{max,out}$. Both findings are intuitive. Measures that do almost not correlate are fill $p$, reciprocity $y$, the pseudo-diameter $\delta$, and the power-law-exponent $\alpha$ (cf. Figure \ref{fig:answerdist}). Hence, regardless of the group of measures and use case of interest, we conclude that the following minimal set of graph measures can be considered in order to characterize an RDF dataset: $n$, $m$, $d_{max}$, $z$, fill $p$, reciprocity $y$, pseudo-diameter $\delta$, and the power-law-exponent $\alpha$.

\section{Conclusions and Future Work}
\label{sec:conclusion}

In this paper, we first introduce a software framework to acquire and prepare RDF datasets. By this means, one can conduct recurrent, systematical, and efficient analyses on their graph topologies. Second, we provide the results of the analysis conducted on 280 datasets from the LOD Cloud 2017 together with the datasets prepared by our framework. 
We have motivated our work by mentioning usage scenarios in at least three research areas in the Semantic Web: synthetic dataset generation, graph sampling, and dataset profiling. In a preliminary analysis of the results, we reported on observations in the group of basic graph measures, degree-based measures, and degree distribution statistics. We have found that (1) the average degree across all domains is approximately 8, (2) without regard to some exceptional datasets, the average degree does not depend on the volume of the graphs (number of edges). Furthermore, (3) due to the way how datasets are modelled, there are domain- and dataset-specific phenomena, e.g., an $h$-index that is constant with the size of the graph on one hand, and an exponentially growing $h$-index on the other.

We can think of various activities for future work. We would like to face the question what actually causes domain- and dataset-specific irregularities and derive implications for dataset modelling tasks. Further, we would like to investigate correlation analyses of graph-based measures with measures for quality of RDF datasets or for data-driven tasks like query processing. For this reason, the experiment will be done on a more up-to-date version of datasets in the LOD Cloud. In the next version we are planing to publish a SPARQL endpoint to query datasets and measures from the graph-based analyses.

\vspace{-0.25cm}
\bibliographystyle{splncs04}

\begin{thebibliography}{10}
\providecommand{\url}[1]{\texttt{#1}}
\providecommand{\urlprefix}{URL }
\providecommand{\doi}[1]{https://doi.org/#1}

\bibitem{alstott2014}
Alstott, J., Bullmore, E., Plenz, D.: powerlaw: a python package for analysis
  of heavy-tailed distributions. PloS one  \textbf{9}(1),  e85777 (2014)

\bibitem{bachlechner2007}
Bachlechner, D., Strang, T.: Is the semantic web a small world? In: ITA. pp.
  413--422 (2007)

\bibitem{benellefi2018}
Ben~Ellefi, M., Bellahsene, Z., John, B., Demidova, E., Dietze, S., Szymanski,
  J., Todorov, K.: {RDF Dataset Profiling - a Survey of Features, Methods,
  Vocabularies and Applications}. The Semantic Web Journal  \textbf{9}(5),
  677--705 (2018)

\bibitem{debattista2018}
Debattista, J., Lange, C., Auer, S., Cortis, D.: Evaluating the quality of the
  lod cloud: An empirical investigation. The Semantic Web Journal
  \textbf{9}(6),  859--901 (2018)

\bibitem{demter2012}
Demter, J., Auer, S., Martin, M., Lehmann, J.: {LODStats} - an extensible
  framework for high-performance dataset analytics. In: EKAW (2012)

\bibitem{ding06}
Ding, L., Finin, T.: Characterizing the semantic web on the web. In:
  International Semantic Web Conference, ISWC (2006)

\bibitem{duan2011}
Duan, S., Kementsietsidis, A., Srinivas, K., Udrea, O.: Apples and oranges: a
  comparison of {RDF} benchmarks and real {RDF} datasets. In: ACM SIGMOD. pp.
  145--156. ACM (2011)

\bibitem{fernandez2018}
Fern\'andez, J.D., Mart\'inez-Prieto, M.A., de~la Fuente~Redondo, P.,
  Guti\'errez, C.: Characterising {RDF} data sets. JIS  \textbf{44}(2),
  203--229 (2018)

\bibitem{flores}
Flores, A., Vidal, M., Palma, G.: Graphium chrysalis: Exploiting graph database
  engines to analyze {RDF} graphs. In: {ESWC} Satellite Events. pp. 326--331
  (2014)

\bibitem{freeman1979}
Freeman, L.C.: Centrality in social networks: Conceptual clarification. Social
  Networks  \textbf{1}(3),  215--239 (1979)

\bibitem{hirsch2005}
Hirsch, J.E.: An index to quantify an individual's scientific research output.
  Proc. {National Academy of Sciences of the United States of America}
  \textbf{102}(46) (2005)

\bibitem{hogan2010}
Hogan, A., Harth, A., Passant, A., Decker, S., Polleres, A.: Weaving the
  pedantic web. In: {LDOW} (2010)

\bibitem{leskovec2006}
Leskovec, J., Faloutsos, C.: Sampling from large graphs. In: SIGKDD. pp.
  631--636 (2006)

\bibitem{mihindukulasooriya2015}
Mihindukulasooriya, N., Poveda{-}Villal{\'{o}}n, M., Garc{\'{\i}}a{-}Castro,
  R., G{\'{o}}mez{-}P{\'{e}}rez, A.: Loupe - an online tool for inspecting
  datasets in the linked data cloud. In: {ISWC} Posters {\&} Demonstrations.
  (2015)

\bibitem{newman2010}
Newman, M.E.J.: Networks: An Introduction. Oxford University Press (2010)

\bibitem{page1999}
Page, L., Brin, S., Motwani, R., Winograd, T.: The pagerank citation ranking:
  Bringing order to the web. Tech. rep., Stanford InfoLab (1999)

\bibitem{qiao2015}
Qiao, S., Özsoyoglu, Z.M.: {RBench: Application-Specific RDF Benchmarking}.
  In: SIGMOD. pp. 1825--1838. ACM (2015)

\bibitem{schmachtenberg2014}
Schmachtenberg, M., Bizer, C., Paulheim, H.: Adoption of the linked data best
  practices in different topical domains. In: ISWC. pp. 245--260 (2014)

\bibitem{sejdiu2018}
Sejdiu, G., Ermilov, I., Lehmann, J., Mami, M.N.: {DistLODStats}: Distributed
  computation of {RDF} dataset statistics. In: ISWC. pp. 206--222 (2018)

\bibitem{tay2011}
Tay, Y.C.: Data generation for application-specific benchmarking. PVLDB,
  Challenges and Visions  \textbf{4}(12),  1470--1473 (2011)

\bibitem{theoharis2008}
Theoharis, Y., Tzitzikas, Y., Kotzinos, D., Christophides, V.: On graph
  features of semantic web schemas. {IEEE} TKDE  \textbf{20}(5),  692--702
  (2008)

\end{thebibliography}

\end{document}